
\input harvmac
\noblackbox

\def\l{\lambda}
\def\k{\kappa}
\def\a{\alpha}
\def\d{\delta}
\def\t{\tau}
\def\ps{\psi}

\def\meas{\oint {dz\over 2 \pi i z}}
\def\part{\partial}
\def\lb{\lbrack}
\def\rb{\rbrack}
\def\nl{ {N\over\l}  }
\def\nl2{ {N\over {2\l}}  }
\def\zd{ z\partial_z }

\def\None{ N^{-{1\over {2k+1} } } }
\def\Ntwo{ N^{-{2\over {2k+1} } } }

\def\cp{ {\cal P } }
\def\cq{ {\cal Q } }
\def\cqt{ {\tilde {\cal Q }} }
\def\cpt{ {\tilde {\cal P }} }
\def\bp{ {\bf P} }
\def\bpt{ {\bf {\tilde P} } }

\def\R{ {\rm R} }
\def\to{\tau_1}
\def\tt{\tau_2}
\def\pa{\partial}
\def\o{\over}
\def\Gr{Gr^{(0)}}
\def\Grn{Gr(k,N)}
\def\vo{V_1}
\def\vt{V_2}

\def\so{S_1}
\def\st{S_2}

\def\soi{S_1^{-1}}
\def\sti{S_2^{-1}}

\def\psd{\psi^{\dagger}}
\def\ph{\phi}
\def\C{ {\rm C} }
\def\L{ {\rm L} }
\def\B{ {\rm B} }
\def\F{ {\rm F} }
\def\GL{ { {\rm GL}(\infty) } }
\def\gl{ { {\rm gl}(\infty) } }
\def\GLN{ { {\rm GL}(N) } }
\def\gln{ { {\rm gl}(N) } }
\def\Z{ {\bf{\rm Z}} }
\def\Cn{ {\bf{\rm C}}^N }
\def\poo{P_{00}(z)}
\def\pio{P_{10}(z)}
\def\poi{P_{01}(z)}
\def\pii{P_{11}(z)}
\def\ha{{1\over2}}

\Title{\vbox{\baselineskip12pt\hbox{SU-4238-504}}}
{{\vbox{\centerline{Unitary One Matrix Models:}
\vskip2pt \centerline{String Equation and Flows$^*$}
}}}
\bigskip
\centerline{ Konstantinos\footnote{}{$^*$ Invited talk delivered by
M. J. Bowick at the ${\rm V}^{\rm th}$
Regional Conference on Mathematical Physics, Edirne, Turkey;
December 15-22, 1991.} N. Anagnostopoulos\footnote{}
{$^1$E-mail: Konstant@suhep.bitnet; Bowick@suhep.bitnet.}
{\rm and}  $~$Mark J. Bowick$^1$}
\medskip
\centerline{ Physics Department }
\centerline{ Syracuse University}
\centerline{ Syracuse, NY 13244-1130, USA}
\bigskip
\centerline{\bf Abstract}
\medskip
We review the Symmetric Unitary One Matrix Models. In particular we discuss the
string equation in the operator formalism, the mKdV flows and the Virasoro
Constraints. We focus on the $\t$-function formalism for the flows and we
describe its connection to the (big cell of the) Sato Grassmannian $\Gr$ via
the Plucker embedding of $\Gr$ into a fermionic Fock space. Then the space
of solutions
to the string equation is an explicitly computable
subspace of $\Gr\times\Gr$ which is invariant under the
flows.
\noindent
\Date{ February 29, 1992}
\vfill \eject
%
%
%
%
\centerline{\bf 1. Introduction} \bigskip
One Matrix Models are quantum mechanical systems whose partition function is
defined by an integral of the form:
\eqn\partgen{
Z_M=\int\, dM\, \exp\{ {-{N\o\l} \tr V(M)} \}\, ,
}
where $M$ is an $N\times N$ matrix and the potential $V(M)$ is a polynomial in
$M$. In the last few years, there has been tremendous progress \nref\brka{
Br\'ezin, E. and  Kazakov, V.:
Exactly Solvable Field Theories of Closed Strings. Phys. Lett. {\bf B236},
144-149 (1990).}\nref\dosh{Douglas, M. and  Shenker, S.:
Strings in Less Than One Dimension.  Nucl. Phys. {\bf B335},  635-654
(1990).}\nref\grmi{ Gross, D. and  Migdal, A.:
Nonperturbative Two-Dimensional Quantum Gravity. Phys. Rev. Lett. {\bf 64},
 127-130 (1990).}\nref\grmit{ Gross, D. and  Migdal, A.:
A Nonperturbative Treatment of Two-Dimensional Quantum Gravity. Nucl. Phys.
 {\bf B340},  333-365 (1990).}\refs{\brka{--}\grmit}
in matrix models through the
discovery of a connection of a certain class of these models to two dimensional
gravity coupled to $(p,q)$
minimal conformal matter. This happens when $M$
is a hermitian matrix (HMM) or when one considers generalizations to a $(q-1)$
hermitian multi-matrix model (MHMM), where $(q-1)$ hermitian matrices are
coupled linearly to each other \nref\doug{Douglas, M. R.:
Strings in Less Than One Dimensions and the Generalized KdV Hierarchies.
Phys. Lett. {\bf B238},  176-180 (1990).}\doug .
In the
former simpler case,
the Feynman graphs of the zero dimensional field theory are viewed
as being dual to a discrete dynamical polygonation of an oriented two
dimensional Riemann surface. Then the perturbation series can be summed in the
form
\eqn\partm{
Z_M=\sum_{h=0}^{\infty} N^{\chi}Z_h
}
where $\chi=V-E+L$ is the Euler character of the corresponding surface and $h$
is its genus given by $\chi=2-2h$. Since the number of vertices $V$, edges $E$
and loops $L$ of the Feynman graph correspond respectively
to the number of faces $F$, edges
$E$ and vertices $V$ of the dual graph, the above series can be
shown to correspond to the discretized version of the partition function of
pure two dimensional gravity
\eqn\partgr{
Z_{Gra}=\sum_h \sum_T {1\o C(T)} \exp\biggl({-\mu_B A+{1\o 4\pi G_B}\chi}
\biggr) \, .
}
In \partgr\ $A$ is the area of the surface, $\mu_B$ and $G_B$ the bare
cosmological and Newton's constant and $C(T)$ is the symmetry factor of the
polygonation corresponding to dividing by the volume of the diffeomorphism
group of the surface. The equality of \partm\ and \partgr\ is achieved by
identifying $\l={\rm e}^{-\mu_B}$ and $N={\rm e}^{1\o 4\pi G_B}$. The
action in \partgr\
can also be viewed as the action
of a string theory embedded in  zero dimensional
spacetime
\eqn\strac{
S_{str}=\log{\k_B}\int d^2\xi\sqrt{g}R+\mu_B\int d^2\xi\sqrt{g}\, .
}
Then \partm\ gives the genus perturbation expansion with
$\k_B={1\o N}$, the bare string coupling. The naive continuum theory is taken
by letting $N\rightarrow\infty$. In this case the area $A$ diverges and the
polygonated surface is thought to approach a smooth Riemann surface. For a
critical value $\mu_c$ of $\mu_B$ the increasing
entropy of large surfaces compensates the Boltzmann factor and the system
undergoes a (third order) phase transition. If the
critical point is approached
in an arbitrary way, only the sphere $Z_0$ contributes to
\partm . The remarkable observation \refs{\brka{--}\grmit}
was that since the singular part of
$Z_h\sim (\mu_B-\mu_c)^{\chi (1+{1\o 2k}) }$ with $k$ a positive
integer, one can
obtain contributions from all genera by simultaneously taking
the large $N$-limit and letting $\mu_B$
approach its critical value $\mu_c$ in a coordinated way. The integer $k$
labels a series of multicritical points reached by tuning $k$ parameters in the
potential $V(M)$. Introducing a cutoff $a$ in the theory, we define the string
coupling $\k_0$ and renormalized cosmological constant $\mu_R$ to be
\eqn\renoc{
\k_0={ a^{-(2+{1\o k}) } \o N}\, ,\qquad \mu_R={\mu_B-\mu_c \o a^2}\, .
}
The double scaling limit is defined by taking $N\rightarrow\infty$ and
$\mu_B\rightarrow\mu_c$ while
keeping $\k_0$ and $\mu_R$ fixed. Then the continuum
limit of \partm\ becomes
\eqn\partsc{
Z_{str}=\sum^\infty_{h=0} \k^\chi Z_h\, ,
}
with $\k={\k_0\o\mu_R}$. The series \partsc\ is horribly divergent. It is
non-Borel summable since every term increases as $(2h)!$. This reflects our
ignorance in summing
the perturbation series of string theory although the fixed
genus partition function $Z_h$ can be calculated and is well defined. Happily,
the theory is exactly solvable at the multicritical points
and its dynamical content is given by a single differential equation, the
string equation. The string equation is a differential equation in the variable
$x$
satisfied by the
specific heat $-\pa^2 \log{Z}$, with
$\k^2=x^{ -(2+{1\o k}) }$. It possesses solutions that in the weak coupling
limit $\k\rightarrow 0$ are asymptotic to \partsc\ and we say that the double
scaling limit provides a non-perturbative definition of $Z_{str}$. Indeed
comparison with calculations directly from the continuum theory indicates that
$Z_{str}$ corresponds to two dimensional gravity coupled to $(2k-1,2)$ minimal
conformal matter. Even more interesting is the discovery that the double
scaling
limit of $(q-1)$ MHMM gives two dimensional gravity coupled to $(p,q)$ minimal
conformal matter \doug .

Unitary One Matrix Models (UMM) form another interesting class of matrix
models.
These are defined by \partgen\ with $M$ being a unitary matrix $U$. The
interest
in those models arose a long time ago when Gross and Witten \nref\gros{ Gross,
D. and  Witten, E.: Possible
Third-Order Phase Transition in the Large-N Lattice Gauge Theory.
 Phys. Rev. {\bf D21},  446-453 (1980).}\gros\ showed that the
partition function of two dimensional $U(N)$ QCD on a lattice is given by
$Z_{QCD}=(Z_U)^{V\o a^2}$ and that the theory undergoes a third order phase
transition in the large $N$ limit (V is the volume of the two dimensional world
and $a$ is the lattice cutoff). The theory was also shown to posses a double
scaling limit $N\rightarrow\infty$ and $\l\rightarrow\l_c$ with
$t=(1-{n\o N})N^{2k\o 2k+1}$ and $y=(1-{\l\o\l_c})N^{2k\o 2k+1}$ held fixed
\nref\peri{
Periwal, V.  and Shevitz, D.:
Unitary Matrix Models as Exactly Solvable String Theories.
Phys. Rev. Lett. {\bf 64},  1326-1329 (1990).}\nref\shev{Periwal, V.  and
Shevitz, D.:
Exactly Solvable Unitary Matrix Models: Multicritical Potentials and
Correlations. Nucl. Phys. {\bf B344},  731-746 (1990).}\refs{\peri ,\shev}.
The string equation is a $2k^{\rm th}$ order differential equation of the
function $v$  in the variable $x=t+y$, with $v^2=-\pa^2 \log{Z}$. It has
solutions that are asymptotic to \partsc\ in the
limit $x\rightarrow\infty$ with $\k^2=x^{-(2+{1\o k})}$.
The identifications of those solutions with
conformal field theories coupled to two dimensional gravity or other
interesting systems is still, however, an interesting open problem. Some
interesting suggestions have been made in \nref\cdm{Crnkovi\'c, \v C.,
Douglas, M.  and Moore, G.:
Loop Equations and the Topological Structure of Multi-Cut Models.
Preprint YCTP-P25-91 and RU-91-36.}\cdm . Moreover, the surface
interpretation of UMM is not as clear as in the case of HMM. In \nref\neub{
Neuberger, H.:
Scaling Regime at the Large-N Phase Transition of Two-Dimensional Pure Gauge
Theories. Nucl. Phys.
{\bf B340},  703-720 (1990).}\neub\ Neuberger views
the unitary matrix as $U={\rm e}^{iM}$ where $M$ is hermitian and introduces
$N\times N$ hermitian fermionic matrices $\psi$ and $\overline{\psi}$ to
exponentiate the Haar measure $dU\rightarrow dM det({\delta U\o \delta M})$.
The
resulting surfaces contain an infinite number of types of bosonic vertices
forming bosonic ``webs" and fermionic loops forming their boundaries that might
allow a stringy interpretation of the UMM. For another interesting suggestion
see \nref\mina{Minahan, J. A.: Matrix Models With
Boundary Terms and the Generalized
Painlev\'e II Equation. Phys. Lett. {\bf B268} 29-34 (1991).}\mina .
It is also interesting to note that UMM belong to the same universality class
as the HMM in a different class of multicritical points,
the double-cut HMM \nref\crnkb{Crnkovi\'c, \v C.  and Moore, G.:
Multicritical Multi-Cut Matrix Models.
Phys. Lett. {\bf B257},  322-328 (1991).}\refs{\cdm ,\crnkb}.
This is expected since the critical behaviour is governed by the scaling of
the density of the eigenvalues at the edge of its support  \nref\brbo{
Bowick, M. J.  and Br\'ezin, E.:
Universal Scaling of the Tail of the Density of Eigenvalues in Random Matrix
Models. Phys. Lett.
{\bf B268},  21-28 (1991).}\brbo\ and the eigenvalues
of the two models models scale identically there.

The continuum theory obtained in the double scaling limit has a very rich
mathematical structure. When one considers perturbations by the scaling
operators $<\sigma_k>$ with sources $t_k$, the dependence of the specific
heat (or its square root for the UMM) on the ``times" $t_k$ is given by KdV
flows \nref\bdss{Banks, T., Douglas, M. R., Seiberg, N. and Shenker, S. H.:
Microscopic and Macroscopic Loops in Nonperturbative Two Dimensional Quantum
Gravity.
Phys. Lett. {\bf B238},  279 (1990).}\refs{\doug ,\bdss}
for the HMM and mKdV flows for the (symmetric) UMM. The partition
functions of the theory are found to be given by the corresponding
$\t$-functions \nref\fknc{Fukuma, M.
Kawai, H.  and Nakayama, R.:
Continuum Schwinger-Dyson Equations and Universal Structures in Two-Dimensional
Quantum Gravity.
Int. J. Mod. Phys.
{\bf A6},  1385-1406 (1991).}\nref\dvv{
Dijkgraaf, R., Verlinde,  H.  and Verlinde, E.:
Loop Equations and Virasoro Constraints in Non-Perturbative Two-Dimensional
Quantum Gravity. Nucl. Phys.
{\bf B348},  435-456 (1991).}\nref\holo{
Hollowood, T., Miramontes, L., Pasquinucci, A.  and Nappi, C.:
Hermitian vs. Anti-Hermitian 1-Matrix Models and their Hierarchies.
Preprint IASSNS-HEP-91/59 and PUPT-1280.}\refs{\fknc {--}\holo}
which can be thought as sections of a line bundle over the
Universal Grassmannian. Furthermore the $\t$-functions
that solve the string equation
are annihilated by constraints which for the
one matrix model are the Virasoro constraints \nref\crnka
{Crnkovi\'c, \v C., Douglas, M.  and Moore, G.:
Physical Solutions for Unitary Matrix Models. Nucl. Phys. {\bf B360},
 507-523 (1991).}\refs{\fknc {--}\crnka}. All of those results have
counterparts in the discrete theory. The integrable flows are now with
respect to the couplings in the potential $V(M)$. For the UMM these are given
by Toda flows on the half line \nref\bowi{Bowick,
M.~J., Morozov, A. and
Shevitz, D.: Reduced Unitary Matrix Models and the Hierarchy of $\t$-Functions.
Nucl. Phys. {\bf B354},  496-530 (1991).}\bowi\
and the partition function is given by the
product of two Toda-chain $\t$-functions. The Virasoro constraints $L_n$
have the
simple interpretation of corresponding to invariance of the partition function
under specific transformations, which for the UMM are given by
$\delta U=\epsilon_n(U^{n+1}-U^{1-n})$.

An interesting observation is that the string equation can be written in the
form $\lb P,Q \rb=1$ where $P$ and $Q$ are differential
operators for the HMM \doug\
and $2\times 2$ matrices of differential operators of specific order for the
UMM \nref\abi{Anagnostopoulos, K. N.,
Bowick, M. J.  and Ishibashi, N.:
An Operator Formalism for Unitary Matrix Models.
Mod. Phys. Lett. {\bf A6},  2727-2739 (1991).}\abi .
They correspond to the continuum limits of operators
acting on the space of orthonormal functions used to solve the model. One can
use this form of the string equation to determine easily the points in the
Universal Grassmannian that solve the string equation \nref\aschw{
Schwarz, A.:
On Solutions to the String Equation.
Mod. Phys. Lett. {\bf A6},  2713-2725 (1991).}\aschw .
For the UMM \nref\abs{ Anagnostopoulos K. A.,
Bowick M. J., Schwarz A. S.: The Solution Space of the
Unitary Matrix Model String Equation and the Sato Grassmannian. Preprint
SU-4238-497 (to appear in Commun. Math. Phys.).}
these are found to correspond to a pair of points $\vo$ and $\vt$ in the
(big cell of the) Sato Grassmannian satisfying certain invariance conditions.
It is very
important that the mKdV evolution of $\vo$ and $\vt$ gives new solutions to the
string equation. The $\t$-functions that correspond to $\vo$ and $\vt$ are
shown to satisfy the Virasoro constraints in this formalism \abs\ since the
constraints are derived from the same invariance conditions that solutions to
the string equation satisfy \nref\kasch{
Kac, V.  and Schwarz, A.:
Geometric Interpretation of the Partition Function of 2D Gravity.
Phys. Lett. {\bf B257}, 329-334 (1991).}\nref\fknb{
Fukuma, M., Kawai,  H.  and Nakayama, R.: Infinite-Dimensional
Grassmannian Structure of Two-Dimensional Gravity.
Commun. Math. Phys. {\bf 143} 371-403 (1992).}\nref\schwa{Schwarz, A.:
On Some Mathematical Problems of 2D-Gravity and $W_h$-Gravity.
Mod. Phys. Lett. {\bf A6},  611-616 (1991).}\nref\fkna{
Fukuma, M., Kawai,  H. and Nakayama, R.: Explicit Solution for $p-q$
Duality in Two Dimensional Gravity. Preprint
UT-582-TOKYO.}\refs{\kasch {--}\fkna}.

This  article is organized as follows. In section 2 we review the discrete
formulation of the symmetric UMM. The method of the
orthogonal polynomials in the trigonometric basis is
summarized and the Toda flows and Virasoro constraints are discussed. In
section 3 we describe the double scaling limit and describe how the mKdV flows
arise. In section 4 we give a non-rigorous approach to the connection between
the Sato Grassmannian and the mKdV flows starting from the finite dimensional
Grassmannians. In section 5 we describe the connection of the Sato Grassmannian
to the solutions to the string equation.
\bigskip
\centerline{\bf 2. The Symmetric Unitary Matrix Model} \bigskip
In this paper we will study the UMM defined by the
one matrix integral
\eqn\partuni{
Z_N^U= \int DU\, {\rm exp}\{ -{N\over \l}\, \Tr \, V(U+U^{\dagger}) \}\, ,
}
where $U$ is a $2N\times 2N$ or a $(2N+1)\times (2N+1) $ unitary matrix,
 $DU$ is the Haar measure for the unitary group and the potential
\eqn\pote{
V(U)=\sum_{k\geq 0}\, g_k\, U^k \, ,
}
is a polynomial in $U$.
As standard
we first reduce the above integral to an
integral over the eigenvalues \nref\brez{Br\'ezin, E.
Itzykson, C., Parisi,  G.  and Zuber, J.B.:
Planar Diagrams.
Commun. Math. Phys. {\bf 59},  35-51 (1978).}\refs{\gros , \brez}
$z_i$ of $U$ which lie
on the unit circle in the complex $z$ plane.
\eqn\partunit{
Z_N^U= \int \lbrace \prod_j
{ {dz_j}\over {2\pi i z_j} } \rbrace
\, | \Delta (z)|^2 {\rm exp}
\{ -{N\over \l}\, \sum_i V(z_i+z^*_i)\} \, ,
}
where $\Delta (z)= \prod\limits_{k < j}\, (z_k-z_j)$
is the Vandermonde determinant. The Vandermonde determinant is conveniently
expressed in terms of trigonometric
orthogonal polynomials \nref\myer{
Myers, R.C.  and Periwal, V.:
Exact Solutions of Critical Self Dual Unitary Matrix Models
Phys. Rev. Lett. {\bf 65},  1088-1091
(1990).}\myer\
\eqn\poly{
 \eqalign{
    c_n^{\pm}(z)
         &=z^n\pm z^{-n}+
        \sum_{i=1}^{i_{max}}\alpha^{\pm}_{n,n-i}(\,z^{n-i}\pm\,z^{-n+i})\cr
         &=\pm c_n^{\pm}(z^{-1})\cr
          }
}
where for $U(2N+1)$ $n$ is a non-negative integer and $i_{max}=n$
and for U(2N) $n$ is
a positive half-integer and $i_{max}=n-{1\o 2}$ . The polynomials
$c_n^{\pm}(z)$ are orthogonal with respect to the inner product
\eqn\inner{
\eqalign {
  \langle c_n^{+},c_m^{+}\rangle
     &= \meas\, \exp \{ -{N\over \l}\, V(z+z^*)\}\, c_n^{+}(z)^* c_m^{+}(z)
\cr
     &={\rm e}^{\phi_n^{+}}\,\delta_{n,m} \, ,\cr
  \langle c_n^{-},c_m^{-}\rangle &={\rm e}^{\phi_n^{-}}\,\delta_{n,m} \, ,\cr
  \langle c_n^{+},c_m^{-}\rangle &=0 \, .\cr
          }
}
The expression for the Vandermonde determinant is
\eqn\vanpoe{
|\Delta(z)|^2=\bigg| det\pmatrix{c_i^-(z_j)\cr c_i^+(z_j)\cr}  \bigg|^2
\quad ,
}
where $j=1,\ldots ,2N$,
$i={1\o 2},{3\o 2},\ldots ,N-{1\o 2}$ for $U(2N)$ and
$j=1,\ldots ,2N+1$,
$i=0,1,\ldots,N$ for $U(2N+1)$
(where the line $c_0^-(z)\equiv 0$ is understood to be omitted).
Then the partition function of the model is given by the product of the norms
of the orthogonal polynomials
\eqn\parttau{
Z^U_N=\prod_n {\rm e}^{\phi_n^{+}}{\rm e}^{\phi_n^{-}}=
\tau_N^{(+)}\tau_N^{(-)}\, .
}
The functions $\tau_n^{(+)}$ and $\tau_n^{(-)}$ are Toda chain $\t$-functions
on the half line \bowi\
\eqn\todaeq{
{\pa^2 \phi_n^{\pm}\o \pa g_1^2}={\rm e}^{\phi_{n+1}^\pm-\phi_n^\pm}-
                                 {\rm e}^{\phi_{n}^\pm-\phi_{n-1}^\pm} \, ,
}
with solutions ${\rm e}^{\phi_{n}^\pm}={\tau_{n+1}^{(\pm)}\o \tau_n^{(\pm)} }$.

The orthogonal basis of polynomials chosen is especially useful for
constructing the operator formalism of the theory. When acting on the basis of
orthonormal functions
\eqn\funpi{
\pi_n^{\pm}(z)={\rm e}^{-\phi_n^{\pm} /2 } {\rm e}^{-{N\over {2\l} } V(z_+) }
c_n^{\pm}(z)
}
such that
\eqn\innpi{
\eqalign{
\langle \pi_n^{+}(z),\pi_m^{+}(z) \rangle
                 &=\meas \pi_n^{+}(z)^* \pi_m^{+}(z)\cr
                 &= \delta_{n,m} \, ,\cr
\langle \pi_n^{-}(z),\pi_m^{-}(z) \rangle&= \delta_{n,m} \, ,\cr
\langle \pi_n^{+}(z),\pi_m^{-}(z) \rangle&= 0 \, ,\cr
        }
}
the operators $z_{\pm}=z\pm{1\over z}$ and $\zd$
give finite term recursion relations
%
\eqn\recplp{
\eqalign{
z_+\, \pi_n^{\pm}(z)&=
\sqrt{R_{n+1}^{\pm} }\pi_{n+1}^{\pm}(z)-r_n^{\pm} \pi_{n}^{\pm}(z) +
\sqrt{R_{n}^{\pm} } \pi_{n-1}^{\pm}(z)~,   \cr
z_-\, \pi_n^{\pm}(z)&=
\sqrt{Q_{n+1}^{\mp} }\pi_{n+1}^{\mp}(z)-q_n^{\pm}
\sqrt{Q_n^{\mp}\over R_n^{\pm}  } \pi_{n}^{\mp}(z) -
\sqrt{Q_{n}^{\pm} } \pi_{n-1}^{\mp}(z)   \, ,   \cr
\zd \pi_{n}^{\pm}(z)&=
-\nl2 \sum_{r=1}^k (v_z^{\pm})_{n,n+r} \pi_{n+r}^{\mp}(z)
+ \biggl \{ n \sqrt{ Q_n^{\mp}\over R_n^{\pm}  }-\nl2 (v_z^{\pm})_{n,n} \biggr
\}
\pi_{n}^{\mp}(z) \cr
       &\, +\nl2 \sum_{r=1}^k (v_z^{\pm})_{n,n-r} \pi_{n-r}^{\mp}(z)~~,\cr
           }
}
where $R_n^{\pm}={\rm e}^{\phi_n^{\pm}-\phi_{n-1}^{\pm} }$,
 $Q_n^{\pm}={\rm e}^{\phi_n^{\pm}-\phi_{n-1}^{\mp} }$,
 $r^{\pm}_n={ {\partial \phi_n^{\pm} }\over {\partial g_1} } $,
 $q_n^{\pm}= { { (Q_{n+1}^{\pm}-Q_n^{\pm})+(R_{n+1}^{\mp}-R_n^{\pm}) }\over
              { r_n^{\pm}-r_n^{\mp} }  }$, and

%
%
$$(v_z^{\pm})_{n,n-r}=\meas \pi_{n-r}^{\mp}(z) ^*\, \{ \zd V(z_+)\} \,
   \pi_{n}^{\pm}(z) \, ~.
$$
Then the discrete string equation is given by the relation
$\lb\zd,z_\pm\rb=z_\mp$.

Invariance of the partition function under the transformations
$$\delta_n U=\epsilon (U^{n+1}-U^{1-n})\qquad n\geq 1\, ,$$
implies that the partition function is annihilated by the Virasoro constraints
\eqn\virade{
\L_n=\sum_{k=0}^\infty kg_k{\pa\o \pa g_{k+n}}+ \ha
\sum_{1\leq k\leq n} {\pa^2\o \pa g_k\pa g_{n-k} }\, .
}
In \bowi\ it was argued that the string equation can be viewed as a consistency
condition of the integrable hierarchy and the Virasoro constraints.
\bigskip
\centerline{\bf 3. The Double Scaling Limit}\bigskip
The continuum limit of \partuni\ is taken by letting $N\rightarrow\infty$. Then
the eigenvalues $\a_i$, where $z_i={\rm e}^{i\a_i}$, become continuously
distributed over the unit circle $|z|=1$ and their distribution is described by
the density of eigenvalues
\eqn\dens{
\eqalign{ \rho(\a)={ds\o d\a}\, ,\quad   & s={i\o N}\cr
          \int^{a_c}_{-a_c}\rho(\a)d\a=1 \quad & 0<a_c\leq\pi\, .\cr
        }
}
If $\rho(\a)$ is given, quantities of physical interest, like the free energy
$F_{sph}={1\o N^2}\log{Z}$,  can be calculated. For example the saddle point
approximation of \partuni\ gives
\eqn\freen{
F_{sph}={2\o\l}\int^{a_c}_{-a_c} d\a \rho(\a)V(2\cos{\a})+
        P\,\int^{a_c}_{-a_c}d\a d\beta\rho(\a)\rho(\beta)
        \log{|\sin{\a-\beta\o 2}|}+{\rm const.}\, ,
}
where $P$ denotes the principal value of the integral. Then one can think of
the eigenvalues as a Dyson gas of electric charges on the unit circle subject
to their mutual Coulomb repulsion and an external potential $V$. In the weak
coupling limit $\l\rightarrow\infty$ the eigenvalues tend to distribute
uniformly on the circle, whereas in the strong coupling limit
$\l\rightarrow 0$ the charges are localized, say at the point $z=1$. The system
undergoes a phase transition precisely when the eigenvalue distribution
develops a cut at $z=-1$ and it happens when $\l_c=1$. Near the cut
$\rho(\a)$ scales as
\eqn\evcri{
\rho_k(\a)\sim c_k(1-\sin^2{\a\o2})^k\qquad \a\rightarrow\pi\, ,
}
and we obtain a third order phase transition with
$F\sim(\l-\l_c)^{2+{1\o k}}$ \gros .
The $k^{\rm th}$ multicritical point is obtained
by tuning $k$ couplings in the potential $V(U)$ to their critical values.

The double scaling limit \refs{\peri ,\shev}
corresponding to the $k^{\rm th}$ multicritical
point is defined by $N\rightarrow\infty$ and $\l\rightarrow\l_c$, with
$t=(1-{n\over N})N^{ {2k}\over {2k+1} }$,
$y=(1-{\l\over\l_c})N^{ {2k}\over {2k+1} }$ held fixed.
It was shown in \abi\ that the operators $z_{\pm}$ and $\zd$
have a smooth continuum limit given by
\eqn\contzform{
\eqalign{
z_+ & \rightarrow 2+ \Ntwo \, {\cal Q}_+ \, ,  \quad
z_-\rightarrow - 2\None \, {\cal Q}_- \, ,  \cr
\zd & \rightarrow N^{ {1\over {2k+1} } } ~{\cal P}_k ~,\cr
        }
}
where ${\cal Q}_{\pm}$ are given by
\eqn\contzp{
\eqalign{
{\cal Q}_-&=
\pmatrix{0&\partial  +v\cr
         \partial  -v&0\cr} \, ,    \cr
{\cal Q}_+&=
\pmatrix{ {(\partial+v)(\pa -v) }&{0}\cr
         {0}&{ (\partial -v)(\pa +v) }\cr } \cr
          &={\cal Q}_-^2\, ,\cr
        }
}
and ${\cal P}_k$ by
\eqn\glob{{\cal P}_k=\left(
\matrix{
        0&{\bf P}_k\cr
{\bf P}^{\dagger}_k&0\cr
       } \right) \, .
}
Here
$\partial\equiv { \partial\over {\partial x} }$ and $x=t+y$.
The scaling function
$v^2$ is proportional to the specific heat $-\pa^2\ln Z$ of the model.
The operators $\bp_k$ are differential operators of order $2k$.
The same assertions hold if we introduce sources $t_{2k+1}$($t_1\equiv x$)
and deform the $k^{\rm th}$ multicritical
potential $V_k$ to $V_k(z)-\sum\limits_l t_{2l+1} V_l(z) N^{2(k-l)\o 2k+1}$.
{}From $\lb \zd , z_- \rb=z_+$ it follows that
\eqn\strcon{
\lb {\cp} ,\cq _- \rb =1 \, ,
}
where $\bp=-\sum\limits_{l\geq 1}(2l+1)t_{2l+1}\bpt_l-x$ with
$\bpt_l=\bp_l+x$. The function $v(x)$ becomes a function of $x$ and the times
$\{ t_{2l+1}\}$ and obeys the string equation
\eqn\perstr{
\sum_{l\geq 1}(2l+1)t_{2l+1}{\hat{\cal D}} \R _{l}\lb u \rb =-vx \, .
}
where ${\hat{\cal D}}=\partial + 2v$, $u=v^2-v'$, and
$\R _{k}\lb u \rb $ are the
Gel'fand-Dikii potentials defined through the recursion relation
\eqn\kdvpot{
\pa\, \R _{k+1}\lb u \rb =\biggl( {1\o 4}\pa ^3-{1\o 2}(\pa u+u\pa) \biggr)
\, \R _{k}\lb u \rb \, ,\quad
\R _{0}\lb u \rb ={1\over 2} \, .
}
The dependence of $v$ on the times $\{ t_{2l+1}\}$ is given by the mKdV flows
\eqn\mkdvfl{
{ {\partial v}\over {\partial t_{2k+1} } }=-\partial
{\hat{\cal D}}  \R _{k}\lb u \rb\, .
}
It is very important that \mkdvfl\ is compatible with the string equation. It
can be shown (see also section 5) that solutions to the string equation
flow with \mkdvfl\ to other solutions of \perstr . The $k^{\rm th}$
multicritical point is reached when $t_{2k+1}=-{a_k\o 2k+1}$ and all other
times are zero. In this case the string equation becomes
\eqn\streq{
{\hat{\cal D}} \R _{k}\lb u \rb =a_k vx \, .
}
This is a $2k^{\rm th}$ order differential equation which as
$x\rightarrow\infty$ has asymptotic solutions of the form
\eqn\vasy{
v\sim x^{1\o 2k}\bigl( 1+\sum_{l=1}^\infty v_l x^{-l(2+{1\o k})}\bigr)\, ,
}
which upon the identification $\k^2=x^{-(2+{1\o k})}$ gives the genus
expansion of the specific heat
\eqn\spasy{
v^2\sim x^{1\o k}\bigl( 1+\sum_{h=1}^\infty f_h \k^{2g}\bigr)\, ,
}
where $f_h=2v_h+\sum_{l_1+l_2=h}v_{l_1}v_{l_2}$.

The connection to the $\t$-function formalism of the mKdV hierarchy is shown
by noting that the specific heat $v^2$ can be written
in the form \holo\
\eqn\vtau{
v^2=-\pa ^2\, \log{(\to\tt)}
}
with $\to$ and $\tt$ the $\t$-functions of the mKdV hierarchy \mkdvfl . These
are simply connected to the Miura transformed functions
$u_1=v^2+v'$ and $u_2=v^2-v'$ by $u_i=-2\pa^2\,\log{\t_i}\, ,\, i=1,2$.
Then the partition function is given by
\eqn\miwa{
Z=\to\cdot\tt      \, ,
}
which is the continuum analog of \parttau .

The Virasoro constraints \holo\ are
obtained by first substituting \mkdvfl\ into
\perstr\ and then using \vtau . The result is
\eqn\viraze{
\L_0\t_i=\mu\t_i\, ,
}
with
$\L_0=\sum_{k=0}^\infty(k+\ha)t_{2k+1}{\pa\o \pa t_{2k+1}}+{1\o 16}$ and $\mu$
an arbitrary constant. The flows and the recursion relations relate
$\L_{n+1}$ to $\L_n$ and one obtains
\eqn\viraal{
\L_n\t_i=0\quad\hbox{with}\quad n\geq 1\, ,
}
where
$\L_n=\sum_{k=0}^\infty(k+\ha)t_{2k+1}{\pa\o\pa t_{2(k+n)+1} }+
\ha\sum_{k=1}^{n}{\pa^2\o \pa t_{2k-1}\pa t_{2(n-k)+1} }$.
We will further discuss the Virasoro constraints in section 5.
\bigskip
\centerline{\bf 4. The mKdV Hierarchy and the Sato Grassmannian}\bigskip
As we already mentioned in the introduction, the analysis of the solutions
of the string equation in the Sato Grassmannian $Gr$ depends crucially on the
association of the mKdV $\t$-functions $\to$ and $\tt$ to points $\vo$ and
$\vt$ in the big cell of the Sato Grassmannian $\Gr$. In this section we take a
pedestrian approach to explaining this association and the reader familiar
with the subject might want to skip to the next section. For more rigorous
treatments on the subject see \abs\ and the references therein.

Since the Sato Grassmannian is an infinite dimensional
generalization of finite
dimensional Grassmannians, we start by reviewing the relevant concepts in the
finite dimensional case. For a nice review along
these lines see \nref\mats{Matsuo, Y.: Universal Grassmann Manifold and
Geometrical Structure of Conformal Field Theory on a Riemann Surface.
Preprint UT-523-TOKYO, Dec. 87 (Ph.D. Thesis).}\mats .
The Grassmannian $Gr(k,N)$ consists of all $k$-dimensional linear subspaces of
$\Cn$. A point $V\in\Grn$ is described by a basis $\{ v_i\}$ with
$i=1,\ldots,k$ and a basis of the orthogonal complement of $V$
$\{ w_i\}$ with $i=k+1,\ldots,N$. Then the pair $(v,w)$ specifies a point in
$\Grn$. A pair $(v',w')$, however, gives the same point if
$$ (v',w')=(v,w)\pmatrix{ A&B\cr 0&C\cr}\, .$$
Then
$$\Grn\simeq\GLN /P$$
with $P=\biggl\{ \pmatrix{ A&B\cr 0&C\cr} \biggr\}$. The relation between
$\Grn$ and
fermions is established by considering the $\GLN$ representation on a fermionic
Fock space $\F$ defined by the vacua
\eqn\devac{
|k>=e_1\wedge \ldots \wedge e_k\qquad
<k|=i_{e_k}\ldots i_{e_1}\,\quad <i|k>=\delta_{ik}\, ,
}
where $\{ e_i\}$ is a basis of $\Cn$ and $i_{e_i}(e_j)=\delta_{ij}$ is the
inner product operator. The fermionic operators are defined by
\eqn\deffe{
\psd_i =e_i\wedge|\chi>\qquad
\psd_i =i_{e_i}|\chi>\, ,
}
and satisfy canonical anticommutation relations
\eqn\fercom{
\{\ps_i,\psd_j\}=\d_{ij}\, ,\quad\{\ps_i,\ps_j\}=\{\psd_i,\psd_j\}=0
\, .
}
The vacua $|k>$ carry charge $k$ and $\psd_i(\ps_i)$ create a charge $+1(-1)$.
Then
\eqn\anvac{
\eqalign{
     \psd_i|k>=0\quad &i=1,\ldots,k\qquad \ps_i|k>=0\quad i=k+1,\ldots,N\cr
     <k|\psd_i=0\quad &i=k+1,\ldots,N\qquad <k|\ps_i=0\quad i=1,\ldots,k\, ,\cr
        }
}
The Plucker embedding is defined by assigning to every point $V\in\Grn$ a state
\eqn\plude{
|v>=c\,\,v_1\wedge\ldots\wedge v_k\quad {\rm with}\quad v_i=\sum v_{ij}e_j\, ,
}
where $\{ v_i\}$ is a basis of $V$ and $c$ is an arbitrary constant. A change
of basis $v_i\rightarrow a_{ij}v_j$ corresponds to
$c\rightarrow ({\rm det}\, a)\,\, c$ and the state $|v>$ is well defined. The
condition
\eqn\ansta{
\psd [v_i]|v>=0\quad\forall i\, ,
}
with $\psd[v_i]=\sum v_{ij}\psd_i$ defines equivalently the state $|v>$ up to
the constant $c$.

Then  $a\in \gln$ acts on $\F$ by
\eqn\deact{
{\hat a}|\chi>=\sum \psd_i a_{ij} \ps_j |\chi>\qquad |\chi>\in\F\, ,
}
and on the space of operators on $\F$ by
\eqn\braop{
\lb \ps_i,\hat{a}\rb=\sum_k a_{ik}\ps_k\, ,\quad
\lb \hat{a},\psd_i\rb=\sum_k \psd_k a_{ki}\, .
}
The action of  $g\in \GLN$ is defined by exponentiation of \deact . For example
\eqn\frepg{
{\hat g}  \psd_{i_1}\psd_{i_2}\ldots\ps_{i_1}\ps_{i_2}\ldots |0>=
       (\psd g)_{i_1}(\psd g)_{i_2}\ldots
        (g\ps)_{i_1}(g\ps)_{i_2}\ldots|0>
}
with $(\psd g)_{i}\equiv\psd_jg_{ji}$ and $(g\ps)_i\equiv g_{ij}\ps_j$.
Then a $\gln$ operator $a$ acting on $V\in \Grn$ by
$a\,v=\sum (a_{ij}v_j)e_i$ corresponds to a fermionic operator
${\hat a}=\sum \psd_i a_{ij} \ps_j$. Then if
$\hat{a}_1\leftrightarrow a_1$ and $\hat{a}_2\leftrightarrow a_2$,
equations \braop\ give
\eqn\bracor{
\lb \hat{a}_1, \hat{a}_2\rb \leftrightarrow\lb a_1,a_2\rb\, .
}
Moreover note that if
\eqn\relop{
\hat{a}|v>={\rm const.}|v>\Leftrightarrow a\, V\subset V\, .
}

The state $|v>$ belongs to the $\GLN$ orbit of the state $|k>$. Since for
$|v>=v_1\wedge\ldots\wedge v_k$ every vector $v_i$ can be written in the form
$v_i=g\, e_i$ for some fixed $g\in\GLN$, we have that $|v>={\hat g}|k>$ as
defined in \frepg . Therefore the image of $\Grn$ under the Plucker embedding
can be identified with the orbit $\GLN |k>$.

The $\t$-functions are given by fermion correlators
\eqn\tfcor{
\t_V^{\cal O}=<{\cal O}>_V=<k|{\cal O}|v>\, ,
}
with $\cal O$ a zero charge operator. Since the topology of $\Grn$ is
non-trivial, we divide it into cells $({\cal U}_a, a\in I)$. A point
$V\in {\cal U}_a$ is represented by a basis $\{ v_i^{(a)}\}$ and the state
$|v>^{(a)}=v_1^{(a)}\wedge\ldots\wedge v_k^{(a)}$. Then if
$V\in {\cal U}_a\cap {\cal U}_b$ we have
$v_k^{(a)}=a_{ki}^{(ab)}\,v_i^{(b)}$ and
$$\t_V^{{\cal O}(a)}={\rm det}\, a^{(ab)}\, \t_V^{{\cal O}(b)}$$
Therefore the $\t$-functions are really sections of a determinant line bundle
over $\Grn$ whose transition functions are given by
${\rm det}\, a^{(ab)}$.

Most of the results carry over almost unchanged to the infinite
dimensional case. For the infinite dimensional vector space we consider the
space of formal Laurent series
$$H=\{ \sum\limits_n a_n z^n\, ,\quad a_n=0\quad\hbox{for}\quad n\gg 0\,\}$$
and its decomposition
$$H=H_+ \oplus H_- \, ,$$
where
$H_+=\{ \sum
\limits_{n\geq 0} a_n z^n\, ,\quad a_n=0\quad\hbox{for}\quad n\gg 0\,\}$.
Then the big cell of the Sato Grassmannian $\Gr$ consists of all subspaces
$V\subset H$ comparable to $H_+$, in the sense that the natural projection
$\pi_+ :\, V\rightarrow H_+$ is an isomorphism.
Then $V$ admits a basis of the form $\{\phi_i(z)\}_{i\geq 0}$ where
$\phi_i(z)=z^i+\hbox{lower order terms}$. The Plucker embedding \plude\ is
defined by the semi-infinite wedge product
\eqn\pluinf{
|v>=c\, \phi_1(z)\wedge\phi_2(z)\wedge\ldots\, .
}
Care has to be taken so that a $\GL$ change of basis
$\phi_i(z)\rightarrow a_{ij} \phi_j(z)$ does not introduce infinities, since
${\rm det}\, a$ can be infinite. We choose a set of admissible bases for
$V\in Gr$ to be those whose matrix relating
$\{\pi_+(\phi_i(z))\}_{i\geq 0}$ to $\{ z^i\}_{i\geq 0}$ differs from the
identity by an operator of trace class. Then the fermionic representation
is defined on the Fock space built on the vacuum state of zero charge
\eqn\vacinf{
|0>=\, 1\,\wedge\,z\,\wedge\,z^2\,\wedge\ldots\, ,
}
by fermions $\psd_i$ and $\ps_i$ defined as in \deffe . The states ($m>0$)
\eqn\chgst{
|m>=\psd_m\ldots\psd_1|0>\, ,\quad |-m>=\ps_{-m+1}\ldots\ps_0|0>\,
}
are the filled states with charge $m$ and $-m$ respectively. The generalization
of $\gln$ is given by $\gl$ and is represented on $\F$ by its central extension
$gl^*(\infty)$ with
\eqn\biop{
\hat{a}=\sum_{i,j}:\psd_ia_{ij}\ps_j :
}
where
\eqn\orde{
:\psd_i\ps_j:\,\,=\psd_i\ps_j-<\psd_i\ps_j>=
\cases{\psd_i\ps_j\quad i>0\cr
      -\ps_j\psd_i\quad i\leq 0\cr}
}
is the normal ordering. The reason for introducing normal ordering is that
the naive operator
$\sum_{i,j}\psd_ia_{ij}\ps_j $ maps an admissible basis to a
non-admissible one.

The connection of the fermion representation of $\Gr$ and the KP and mKP
hierarchies is made explicit by making use of the boson-fermion equivalence in
two dimensions. The fermionic currents
\eqn\curdef{
J_n=\sum_{r\in\Z}:\psd_{n-r}\ps_r:\quad n\in\Z
}
satisfy the bosonic commutation relations
\eqn\comcur{
\lb J_m,J_n\rb=m\d_{m,-n}\, .
}
By representing the bosonic Fock space by
$\B \cong\C\lb t_1,t_2,\ldots,;u,u^{-1}\rb$, the space of polynomials in
$t_1,t_2,\ldots,;u,u^{-1}$,
$\pa\o \pa t_n$ and $-nt_{-n}$ (with $n\geq 0$) act as creation and
annihilation operators on $\B$ satisfying the algebra \comcur . Then fermionic
operators can be mapped to operators acting on $\B$ and states in $\F$ to
states in $\B$ by mapping the state $|m>$ of $\F$ to $u^m$. Then the
$k^{\rm th}$ modified KP hierarchy $\t$-functions correspond to correlators
\tfcor\ where ${\cal O}={\rm e}^{\sum_{p\geq 1}t_pJ_p}$ and the states
$|v>$ correspond
to the $\GL$ orbit of $|i>$ with $i=0,\ldots ,k-1$. In particular the
solutions to the second mKP hierarchy is given by two $\t$-functions
\eqn\nstf{
\t_i(t)=<i-1|\exp \{\sum\limits_{p\ge 1}t_pJ_p\}g|i-1>\quad (i=1,2)\, ,
}
where $g\in \GL$.
The modified KdV hierarchy that arises in UMM is the second reduced
mKP hierarchy of the above equation and it corresponds to eliminating from
\nstf\ the dependence on the even times $\{ t_{2n}\}$. Therefore every
solution $\t_1(t)$ and $\t_2(t)$ of the mKdV hierarchy corresponds
to points $\vo(t)$ and $\vt(t)$ in $\Gr$ given by the states
$|v_i(t)>=\exp \{\sum\limits_{p\ge 1}t_pJ_p\}g|i-1>$.
Then the time dependence of $V_i(t)$ is given by
\eqn\feflo{
{\pa\o \pa t_{2k+1}}|v_i(t)>=J_{2k+1}|v_i(t)>\,{\rm and}\quad
J_{2k}|v_i(t)>=0\, ,
}
or by using the correspondence \bracor\
\eqn\spaflo{
{\pa \o\pa t_{2k+1}}\, V_i(t) =z^{2k+1}\, V_i(t)\quad{\rm and}\quad
z^{2k}\, V_i(t)\subset V_i(t)\, .
}
Then $V_i(t)=\exp \{ \sum\limits_k t_{2k+1}z^{2k+1} \}V_i\equiv
\gamma(t,z)V_i$.
\bigskip
\centerline{\bf 5. The Solutions to the String Equation}\bigskip
Since to every solution of the mKdV hierarchy correspond points
$\vo(t)$ and $\vt(t)$ in $\Gr$ satisfying \spaflo , one would like to determine
those that are solutions to the string equation \strcon . This is particularly
easy because the commutator
$\lb {\cp} ,\cq _- \rb$ is equal to a constant \abs .

Consider the space $\Psi$ of pseudodifferential operators
$W=\sum
\limits_{i\leq k}w_i(x)\pa^i$ where the functions $w_i(x)$ are taken to be
formal power series
(i.e. $w_i(x)=
\sum\limits_{k\geq 0}w_{ik}x^{k}\, ,\,\, w_{ik}=0\, ,\,k\gg 0$). $W$ is
then a pseudodifferential operator of order $k$. It is called monic if
$w_k(x)=1$ and normalized if $w_{k-1}(x)=0$. The space $\Psi$ forms an algebra.
The space of monic, zeroth-order pseudodifferential operators forms a group
$\cal G$.

There is a natural action of $\Psi$ on $H$ defined by
$$
\eqalign{
x^m\pa^n : H&\rightarrow H \cr
        \phi&\rightarrow (-{d\o dz})^m(z)^n\, \phi\, .\cr
        }
$$
Then it is well known \nref\mula{The most appropriate exposition for our
purposes is given in Mulase, M.:
Category of Vector Bundles On Algebraic Curves and Infinite Dimensional
Grassmannians. Int. J. Math.
{\bf 1},  293-342 (1990).}\mula\ that
every point $V\in\Gr$ can be uniquely represented in
the form $V=SH_+$ with $S\in{\cal G}$. This will imply that for every operator
$\cq_-$ we can uniquely associate a pair of points $V_1 ,\, V_2\in\Gr$.

Indeed, consider $\so$ and $\st \in{\cal G }$ such that
\eqn\diagq{
{\hat S}\cq_-{\hat S^{-1}}=\cqt_-
}
where
\eqn\diagm{
{\hat S}=\pmatrix{\so&0\cr 0&\st\cr}\, ,\,
\cqt_-=\pmatrix{0&\pa\cr \pa&0\cr}\, .
}
Then
\eqn\condd{
\eqalign{
\so(\pa+v)\sti&=\pa\, ,\cr
\st(\pa-v)\soi&=\pa\, .\cr
        }
}
$\so$ and $\st$ can be shown to exist and are unique up to a redefinition
$S_i\rightarrow S_i\, R$ with
$R=1+\sum_{i\geq 0} r_i\pa^{-i}$ and $r_i$ constants.

Since $V\subset\Gr$ is given uniquely by $V=SH_+$, the operator $\cq_-$
determines two spaces $\vo=\so H_+$ and  $\vt=\st H_+$. Conversely given
spaces $\vo$ and $\vt$ determine
$\cq_-$ uniquely. The operator $\cq_-$, however,
is a differential operator  and $\vo,\vt$
cannot be arbitrary. Indeed, since every differential operator leaves $H_+$
invariant, we obtain
\eqn\incond{
\eqalign{
(\pa+v)\,H_+\subset H_+
\Leftrightarrow&\soi\pa\st\, H_+\subset H_+\cr
\Leftrightarrow&\pa\, \vt\subset\vo\cr
\Leftrightarrow&z\, \vt\subset\vo\cr
        }
}
Similarly, $z\, \vo\subset\vt$. Notice that these conditions are consistent
with the second equation in \spaflo .

The transformation ${\hat S}\cq_-{\hat S^{-1}}=\cqt_-$ is a similarity
transformation and the string equation will be left invariant if we define
$\cpt_{(k)}={\hat S}\cp_{(k)}{\hat S^{-1}}$. Then the solution to
$\lb \cpt_{(k)} ,\cqt_- \rb =1$ is easily found to be given by
\eqn\fourp{
\cpt_{(k)}=\pmatrix{0&A_k\cr A_k&0\cr}\, ,\hbox{\rm where}\quad
A_k={d\o dz}+\sum\limits_{i=0}^k \a_i z^{2i}
\quad\hbox{\rm and $\a_i=$const.}
}

The requirement that $\cp$ be a differential operator
is equivalent to the conditions
$A_k\,\vo\subset\vt$ and $A_k\,\vt\subset\vo$.
The space of solutions
to the string equation is the space of operators $\cq_-$ such that
there exists $\cp_{(k)}$ with $\lb \cp_{(k)} ,\cq_- \rb=1$.
We conclude that this space is isomorphic to the
set of elements $\vo ,\vt\subset\Gr$ that satisfy the conditions:
\eqn\fincon{
\eqalign{
z\,\vo\subset\vt \quad z\,\vt\subset\vo\cr
A_k\,\vo\subset\vt \quad A_k\,\vt\subset\vo\cr
        }
}
for some $A_k={d\o dz}+\sum\limits\limits_{i=0}^k \a_i z^{2i}$.

The string equation is left invariant by the flows \spaflo . Indeed
\eqn\cons{
\eqalign{
z\,\gamma(z,t)V_1\subset\gamma(t,z)V_2&
       \Rightarrow z\,V_1(t)\subset V_2(t)\cr
A_k(t)\,\gamma(z,t)V_1\subset\gamma(t,z)V_2&
       \Rightarrow A_k(t)\,V_1(t)\subset V_2(t)\, ,\cr
        }
}
where
\eqn\newop{
A_k(t)\equiv\gamma A_k \gamma^{-1}=A_k-\sum\limits_k(2k+1)t_{2k+1}z^{2k}
}
and analogous equations with $V_1$ and $V_2$ interchanged.

It is now easy to see that \fincon\ implies the Virasoro constraints for
the $\t$-functions. Without going into the details (see \kasch ),
we first notice
that the operators $l_n=z^{2n+1}\, A$ leave $V_i$ invariant
\eqn\ivir{
z^{2n+1}\, A\,V_i\subset V_i\, .
}
Then using the correspondence \bracor , one can construct the corresponding
fermion operators ${\hat l}_n$ and from them their bosonic counterparts
$\L_n$ \kasch . These have the exact form as equations {\viraze {--}\viraal}.
We can immediately
see that they form a Virasoro algebra by noting that
$l_n\sim z^{2n+1}{d\o dz}$ are the generators of the
Virasoro algebra and by using
lemma \bracor . Since $l_n$ leave $V_i$ invariant, then using \relop\ we
conclude that the operators $L_n$ annihilate the $\t$-functions $\to$ and
$\tt$ and obtain equations \viraze\ and \viraal .

We conclude this section by showing how conditions \fincon\ can be used to
calculate the space of solutions to the string equation \abs .
We will start by describing the spaces
$\vo,\vt$.

First choose vectors $\ph_1(z),\ph_2(z) \in\vo$, such that
$$
\ph_1(z)=1+\hbox{lower order terms}\, ,\quad
\phi_2(z)=z+\hbox{lower order terms}\, .
$$
Then the condition $z^2\,\vo\subset\vo$ and $\pi_+(\vo)\cong H_+$
shows that we can choose a basis for $\vo$
$$
\ph_1,\ph_2,z^2\ph_1,z^2\ph_2,\ldots
$$
Since $z\,\vo\subset\vt$
and $\pi_+(\vt)\cong H_+$ we can choose a basis for $\vt$ to be
$$
\ps,z\ph_1,z\ph_2,z^3\ph_1,z^3\ph_2,\ldots
$$
where $\ps (z)=1+\hbox{lower order terms}$. Using $z\,\vt\subset\vo$ we have
$z\ps=\alpha \ph_1+\beta\ph_2$. Choose $\ph_1 ,\ph_2$ such that
$z\ps=\ph_2$. Then we obtain the following basis for $\vo,\vt$
($\ph\equiv\ph_1$):
\eqn\basis{
\eqalign{
\vo \,:\quad &\ph,z\ps,z^2\ph,z^3\ps,\ldots\cr
\vt \,:\quad &\ps,z\ph,z^2\ps,z^3\ph,\ldots\cr
        }
}
Then it is clear that $\ph,\ps$ specify the spaces $\vo,\vt$. Using the
conditions $A\vo\subset\vt$ and $A\vt\subset\vo$ we obtain
\eqn\stdif{
\eqalign{
({d\o dz}+f_k(z^2))\ph&=\poo\ph+\poi\ps\cr
({d\o dz}+f_k(z^2))\ps&=\pio\ph+\pii\ps \, .\cr
        }
}

Since a generic system of the form \stdif\ will lead to exponential evolution
of the functions $\phi$ and $\ps$, the requirement that they keep their
polynomial form puts severe conditions on $P_{ij}(z)$. A detailed calculation
shows that  the space of solutions to the string equation \strcon\ is the two
fold
covering of the space of matrices $\bigg( P_{ij}(z)\bigg)$ with polynomial
entries in $z$
such that $\poi$ and $\pio$ are even polynomials having equal degree and
leading terms and
$\poo$ and $\pii$ are odd polynomials satisfying the conditions
$\poo+\pii=0$ and ${\rm deg}\poo<{\rm deg}\poi$.


\bigskip\bigskip
\bigskip
\centerline{\bf Acknowledgements}
\bigskip

The research of K.A. and M.B.
was supported by the Outstanding Junior Investigator Grant DOE
DE-FG02-85ER40231, NSF grant PHY 89-04035 and a Syracuse University Fellowship.

\listrefs

\bye